\documentstyle[prl,preprint,aps,epsf]{revtex}
\begin{document}

\title{Nonvanishing Cosmological Constant of Flat Universe in Brane-World 
Scenario} 
\author{
D. K. Park\raisebox{0.8ex}{1,2}\footnote[1]
{Email:dkpark@hep.kyungnam.ac.kr
},
Hungsoo Kim\raisebox{0.8ex}{1}\footnote[2]
{Email:hskim@hep.kyungnam.ac.kr},
S. Tamaryan\raisebox{0.8ex}{3}\footnote[3]
{Email:sayat@moon.yerphi.am}
}
\address{$^1$ Department of Physics, Kyungnam University, Masan, 631-701,
	      Korea   \\
	 $^2$ Michigan Center for Theoretical Physics \\
	 Randall Laboratory, Department of Physics, University of Michigan \\
          Ann Arbor, MI 48109-1120, USA \\
         $^3$ Theory Department, Yerevan Physics Institute,
         Alikhanian Br.St.2, Yerevan-36, 375036,           Armenia}

\maketitle

\date{\today}
\maketitle
\begin{abstract}
The finite temperature effect is examined in Randall-Sundrum brane-world
scenario with inclusion of the matter fields on the brane. At zero
temperature it is found that the theory on the brane is conformally 
invariant, which guarantees $AdS$/CFT. At $4d$ effective action  
we derived a temperature-dependent nonvanishing cosmological constant at
the flat spacetime limit of brane worldvolume. At the cosmological 
temperature $3 {\bf K}$ the cosmological constant is roughly 
$(0.0004 eV)^4$ which is within the upper bound of the recent 
experimental value $(0.01 eV)^4$
\end{abstract}

\newpage
The recent astronomical observations\cite{perl98-1,perl98-2,berna00} indicate 
that our universe is flat and has a nonvanishing positive cosmological
constant. In this letter we will show that these important features of our
universe can be implemented within the Randall-Sundrum(RS) brane-world
scenario\cite{rs99-1,rs99-2} if the effect of nonzero temperature is 
properly involved.

In the RS brane-world scenario our universe is treated as a three 
dimensional brane embedded in the higher dimensional spacetime. This picture 
and its variants are used to solve the various long-standing puzzles 
such as gauge hierarchy\cite{rs99-1}, Newton gravity\cite{rs99-2,garr00}, 
cosmological constant\cite{kim00,alex01}, and black holes\cite{cham99,emp99}.
It is also shown that the brane-world scenario supports a non-static 
cosmological solution\cite{bine00,csa99-1,cli99} which leads to the
conventional Friedmann equation by introducing bulk and brane cosmological
constants and imposing a particular fine-tuning condition between them. 

The RS bulk spacetime is two copies of $AdS_5$ glued in a $Z_2$-symmetric
way along a boundary which is interpreted as the three-brane world-volume.
This fact is very useful to re-interpret the RS scenario within 
$AdS$/CFT or holography principle\cite{verl99}. From the fact that RS
picture yields similar results to those of the Horava-Witten
scenario\cite{hora96} one can imagine that it is somehow related to the 
string theories. In this context it is interesting to examine whether
or not the similarity of RS scenario to $AdS$/CFT is maintained at 
finite temperature. 

It is well-known that $AdS_5$ in $AdS$/CFT is extended to the
Schwarzschild-$AdS_5$\cite{horo91} 
at nonzero temperature by taking a non-extremal limit of 
black 3-brane solution of the string theories. Therefore the 
similarity of RS picture with $AdS$/CFT strongly suggests that the RS 
bulk spacetime
at finite temperature is two copies of the Schwarzschild-$AdS_5$
attached along the boundary as follows
\begin{equation}
\label{naive}
ds^2 = e^{-2k|y|}
\left[ -\left( 1 - \frac{U_T^4}{k^4} e^{4 k |y|} \right) dt^2
       +\sum^3_{i=1} dx^i dx^i \right]
       + \frac{dy^2}{1 - \frac{U_T^4}{k^4} e^{4 k |y|}}
\end{equation}
where 
$k$ is the inverse of $AdS$ radius and 
$U_T$ is the horizon parameter  
proportional to the external temperature.

However, it is shown in Ref.\cite{park01} that the spacetime (\ref{naive})
is not vacuum solution of $5d$ Einstein equation\footnote{Our conventions are 
$M, N = 0, 1, 2, 3, 5$ , $\mu, \nu = 0, 1, 2, 3$ and $i, j=1, 2, 3$.}
\begin{equation}
\label{fieldeq}
R_{MN} - \frac{1}{2} G_{MN} R = 
- \frac{1}{4M^3} \left[\Lambda G_{MN} + v_b G_{\mu \nu}
\delta_M^{\mu} \delta_N^{\nu} \delta(y) \right]
\end{equation}
where $\Lambda$, $M$, and $v_b$ are $5d$ cosmological constant, $5d$ Planck
scale, and brane tension, although we adopt appropriate fine-tuning
conditions. 
This means the similarity of RS scenario to $AdS$/CFT is not trivially
extended to nonzero temperature case.

It is worthwhile noting that Schwarzschild-$AdS_5$ spacetime (\ref{naive})
solves $5d$ Einstein equation (\ref{fieldeq}) in the whole bulk except
only $y = 0$ if one chooses $\Lambda = -24 M^3 k^2$ and 
$v_b = 24 M^3 k (1 - U_T^4 / k^4)$. 
Thus, there exists a possibility to make the spacetime (\ref{naive}) to be
a solution of $5d$ Einstein equation if the content on the brane is 
changed. We will show in the following that it is indeed the case when the 
matter fields are involved on the brane. Furthermore, inclusion of the
matters on the brane naturally provides a nonvanishing cosmological 
constant at $4d$ effective action level even if the world-volume of 
three-brane is flat spacetime.

Considering the matter fields on the brane modifies $5d$ Einstein
equation as follows:
\begin{equation}
\label{modify}
R_{MN} - \frac{1}{2} G_{MN} R =
- \frac{1}{4M^3} \left[\Lambda G_{MN} + (v_b G_{\mu \nu} - S_{\mu \nu})
\delta_M^{\mu} \delta_N^{\nu} \delta(y) \right]
\end{equation}
which is derived by taking a variation to the action
\begin{equation}
\label{modaction}
S = \int d^4 x \int dy \sqrt{-G} 
\left[-\Lambda + 2 M^3 R + (-v_b + {\cal L}_m) \delta(y) \right]
\end{equation}
where ${\cal L}_m$ is Lagrangian for matter fields and 
$S_{\mu \nu}$ is corresponding energy-momentum tensor derived from
${\cal L}_m$;
\begin{equation}
\label{emdefinition}
S_{\mu \nu} = {\cal L}_m G_{\mu \nu} - 2
\frac{\delta {\cal L}_m}{\delta G^{\mu \nu}}.
\end{equation} 
Inserting an {\it ansatz}
\begin{equation}
\label{ansatz1}
ds^2 = e^{-2 \sigma(y)} 
\left[-f(y) dt^2 + \delta_{ij} dx^i dx^j \right] + 
\frac{dy^2}{f(y)}
\end{equation}
into Eq.(\ref{modify}) Einstein equation provides the following three linear 
independent equations;
\begin{eqnarray}
\label{indep1}
6 f \sigma'^{2} - \frac{3}{2} f' \sigma'&=&- \frac{\Lambda}{4 M^3}
						  \\    \nonumber
3 \sigma'' = \frac{1}{4 M^3 f}& & \left[v_b + \frac{S_{00}}{f} e^{2\sigma}
                                        \right] \delta(y)
						 \\     \nonumber
\frac{1}{2}f'' - 2 f' \sigma'&=& \frac{1}{4M^3}
\left[\frac{S_{00}}{f} + S_{11}\right] e^{2\sigma} \delta(y)
\end{eqnarray}
where we assumed $S_{\mu \nu}$ is a diagonalized tensor with 
$S_{11}=S_{22}=S_{33}$. In fact, this assumption is consistent with the 
rotational symmetry of {\it ansatz} (\ref{ansatz1}) in the brane coordinates
$x^i$.

As mentioned before we would like to derive the conditions for the 
Schwarzschild-$AdS_5$ spacetime, {\it i.e.} $f(y) = 1 - \xi e^{4 k |y|}$
and $\sigma(y) = k |y|$ where $\xi = U_T^4 / k^4$, to be a solution
of Eq.(\ref{indep1}). For the convenience we assume that the stress-energy
tensor $S^M_N$ is $S^M_N = \delta(y) \sqrt{1 - \xi}$diag$(-\rho_0, p_0, p_0, 
p_0, 0)$ where $\rho_0$ and $p_0$ are the energy density and pressure of
matters
respectively. For the homogeneous property of our universe we assume that 
$\rho_0$ and $p_0$ are not dependent on the brane coordinate $x^i$. Since
we are considering only the static case now, we just treat $\rho_0$ and $p_0$
as some constants. When, however, we consider a non-static case later, these
will be time-dependent quantities. 

Inserting $S^M_N$ into Eq.(\ref{indep1}) Einstein equations become in the
form;
\begin{eqnarray}
\label{indep-2}
6 f \sigma'^{2} - \frac{3}{2} f' \sigma'&=&- \frac{\Lambda}{4 M^3}
						  \\    \nonumber
3 \sigma'' = \frac{1}{4 M^3 f}& & \left[v_b + \sqrt{1 - \xi} \rho_0\right]
						  \delta(y)
						      \\   \nonumber
\frac{1}{2}f'' - 2 f' \sigma'&=& \frac{\sqrt{1 - \xi}}{4 M^3} (1 + w_0) \rho_0
						  \delta(y)
\end{eqnarray}
where $w_0\equiv p_0/\rho_0$. Hence, it is easy to derive the conditions for
$f(y) = 1 - \xi e^{4 k |y|}$ and $\sigma(y) = k |y|$ to be a solution
of Eq.(\ref{indep-2});
\begin{eqnarray}
\label{fcond}
v_b + \sqrt{1 - \xi} \rho_0&=&24 M^3 k (1 - \xi)   \\   \nonumber
\sqrt{1 - \xi} (1 + w_0) \rho_0&=& -16 M^3 k \xi
\end{eqnarray}
with a fine-tuning condition of $5d$ cosmological constant 
$\Lambda = -24 M^3 k^2$.
Therefore it is impossible to fix $\rho_0$, $p_0$ and $v_b$ completely
within Einstein equation. 

In fact, one can remove the redundancy of the parameters if the $5d$ 
Einstein equation (\ref{modify}) is modified as follows;
\begin{equation}
\label{combi}
R_{MN} - \frac{1}{2} G_{MN} R + \frac{\Lambda}{4 M^3} G_{MN}
= \delta(y) diag(\alpha, \beta, \beta, \beta, 0).
\end{equation}
In this case it is easy to show that the Schwarzschild-$AdS_5$ 
spacetime completely solves Eq.(\ref{combi}) if the fine-tuning
conditions
\begin{equation}
\label{referee2}
\Lambda = -24 M^3 k^2
\hspace{1.0cm}
\alpha = 6k (1 - \xi)^2
\hspace{1.0cm}
\beta = -6k (1 - \frac{\xi}{3})
\end{equation}
are imposed. However, as will be shown later our purpose is to compute the 
induced $4d$ cosmological constant at the effective action level. For this it
is more convenient to treat the density, pressure, and brane tension 
separately.

In order to fix them completely we need to derive 
an additional condition which is linearly independent to conditions 
(\ref{fcond}). Furthermore, the additional condition should fix
$w_0 = 1/3$ at $\xi=0$ to guarantee $AdS$/CFT at zero temperature.
We will show in the following that the additional condition is derived 
if we assume that our finite temperature solution discussed here is 
a static limit of RS
cosmological solution.

RS cosmological solution is non-static solution of Einstein equation 
(\ref{modify}), which is solved by {\it ansatz}
\begin{equation}
\label{ansatz2}
ds^2 = -n^2(t, y) dt^2 + a^2(t, y) \delta_{ij} dx^i dx^j + b^2(t, y) dy^2.
\end{equation}
As shown in Ref.\cite{bine00,csa99-1,cli99} in order to derive a 
Friedmann-type equation it is enough to solve Eq.(\ref{modify}) 
with the {\it ansatz} (\ref{ansatz2}) in the 
vicinity of brane located at $y = 0$. We also introduce time-dependent 
stress-energy tensor on the brane
\begin{equation}
\label{stress}
S^M_N = \frac{\delta(y)}{b} \mbox{diag}(-\rho, p, p, p, 0)
\end{equation}
where $\rho$ and $p$ are assumed to be dependent only on the time due to 
homogeneous property of our universe. Following Ref.\cite{bine00}
we also introduce several notations for convenience:
\begin{eqnarray}
\label{notation}
\Delta f&=& f(0^+) - f(0^-)    \\  \nonumber
\bar{f}&=&\frac{f(0^+) + f(0^-)}{2}   \\   \nonumber
f''&=& \hat{f}'' + \Delta f' \delta(y)
\end{eqnarray}
where $\hat{f}''$ is non-distributional part of $f''$.

Then, it is straightforward to show that the distributional parts of 
$(0,0)$ and $(1,1)$ components of Einstein equation yield
\begin{eqnarray}
\label{0011}
\frac{\Delta a'}{a_0 b_0}&=& - \frac{1}{12 M^3} (\rho + b_0 v_b)
						 \\   \nonumber
\frac{\Delta n'}{n_0 b_0}&=& \frac{1}{12 M^3}
	     \left[(3 p + 2 \rho) - b_0 v_b\right]
\end{eqnarray}
where $a_0$, $b_0$ and $n_0$ are $a$, $b$, and $n$ at $y = 0$
and prime denotes a differentiation with respect to $y$. 
Also it is easy to derive
\begin{equation}
\label{conser}
\dot{\rho} + 3 (\rho + p) \frac{\dot{a_0}}{a_0} + v_b \dot{b_0} = 0
\end{equation}
from $(0,5)$ component of Einstein equation whose explicit expression is 
\begin{equation}
\label{05}
\frac{\dot{a}}{a} \frac{n'}{n} + \frac{a'}{a} \frac{\dot{b}}{b}
- \frac{\dot{a}'}{a} = 0
\end{equation}
where dot denotes a differentiation with respect to $t$.
Eq.(\ref{conser}) is nothing but the conservation condition of the 
stress-energy tensor, {\it i.e.} $S^M_{N;M} = 0$. Finally, the jump of
the $(5,5)$ component of Einstein equation yields 
\begin{equation}
\label{55-1}
\frac{\bar{a}'}{a_0} (p - b_0 v_b) = \frac{\bar{n}'}{3 n_0} (\rho + b_0 v_b)
\end{equation}
and the mean value of the same equation gives the Friedmann-like equation;
\begin{equation}
\label{fried1}
\left(\frac{\dot{a_0}}{a_0}\right)^2 + \frac{\ddot{a_0}}{a_0} = 
\frac{1}{12 M^3} \left[ \Lambda - 
		 \frac{(\rho + b_0 v_b) (\rho + 3 p - 2 b_0 v_b)}{48 M^3}
				    \right].
\end{equation}
Of course, when deriving Eq.(\ref{fried1}) we have used the conditions 
(\ref{conser}), (\ref{05}), and (\ref{55-1}). 

If we assume that our static finite temperature solution is a static 
limit of the RS 
cosmological solution, we can derive the additional condition by taking a 
static limit of the Friedmann-type equation (\ref{fried1});
\begin{equation}
\label{stafried}
(\rho_0 + b_0 v_b) \left[ (1 + 3 w_0) \rho_0 - 2 b_0 v_b \right] = 48 M^3 
\Lambda.
\end{equation}
If one inserts, however, $\Lambda = -24M^3 k^2$ and $b_0 = 1/\sqrt{1 - \xi}$,
Eq.(\ref{stafried}) becomes
\begin{equation}
\label{depen}
\left[\rho_0 + \frac{v_b}{\sqrt{1 - \xi}}\right]
\left[(1 + 3 w_0) \rho_0 - \frac{2 v_b}{\sqrt{1 - \xi}} \right] = -1152 M^6 k^2
\end{equation}
which is not linearly independent to the previous conditions (\ref{fcond}).
Therefore we have to find another linearly independent condition.

As shown in Ref.\cite{csa99-1,cli99} the Friedmann-type equation (\ref{fried1})
is transformed into the conventional Friedmann equation
\begin{equation}
\label{fried2}
\left(\frac{\dot{a_0}}{a_0}\right)^2 + \frac{\ddot{a_0}}{a_0} =
\frac{1}{576 M^6} \left[ v_b b_0 (\rho - 3 p) - \rho (\rho + 3 p) \right]
\end{equation}
when the fine-tuning condition 
\begin{equation}
\label{fine}
\Lambda + \frac{v_b^2 b_0^2}{24 M^3} = 0
\end{equation}
is satisfied. Therefore, 
another candidate for the additional condition is 
either one of 
Eq.(\ref{fine}) or 
static limit of 
Eq.(\ref{fried2}). Both with
the previous conditions (\ref{fcond}) yield identical results;
\begin{eqnarray}
\label{main}
\Lambda&=&-24 M^3 k^2  \\  \nonumber
v_b&=&24 M^3 k \sqrt{1 - \xi}  \\  \nonumber
\rho_0&=& -24 M^3 k (1 - \sqrt{1 - \xi})   \\   \nonumber
w&=& \frac{(3 - \xi) - 3 \sqrt{1 - \xi}}{3 \sqrt{1 - \xi} (1 - \sqrt{1 - \xi})}
\end{eqnarray}
which is our main result in this letter. It is worthwhile noting that
$\lim_{\xi \rightarrow 0} w = 1/3$ which guarantees $AdS$/CFT at zero
temperature. 

Now, let us compute $4d$ effective action to show that the nonzero 
temperature effect makes a nonvanishing cosmological constant at flat
universe by considering a small fluctuation around Schwarzschild-$AdS_5$
\begin{equation}
\label{line1}
ds^2 = e^{-2 \sigma(y)} g_{\mu \nu}(x, y) dx^{\mu} dx^{\nu} +
\frac{dy^2}{f(y)}
\end{equation}
where
\begin{equation}
\label{metric1}
g_{\mu \nu}(x, y) = \bar{g}_{\mu \nu}(x) + (1 - f(y)) \bar{g}_{\mu}^0
						      \bar{g}_{\nu}^0.
\end{equation}
Here, $\bar{g}_{\mu \nu}(x)$ represents a physical gravity in the effective
theory and will be replaced by flat metric $\eta_{\mu \nu}$ at final 
stage. The curvature scalar $R$ computed from the metric (\ref{line1}) is 
\begin{equation}
\label{curvature}
R = e^{2 \sigma} \bar{R} + \Delta R_1 + \Delta R_2,
\end{equation}
where $\bar{R}$ is the $4d$ curvature scalar derived from $\bar{g}_{\mu \nu}$
and
\begin{eqnarray}
\label{remain1}
\Delta R_1&=& 8 f \sigma'' - 20 f \sigma'^2 + 
\frac{\bar{g}^{00} f f'' + [4(1 + \bar{g}^{00}) - 9 f \bar{g}^{00}] f' \sigma'}
     {1 + (1 - f) \bar{g}^{00}}
+ \frac{\bar{g}^{00} (1 + \bar{g}^{00})}
       {2 [1 + (1 - f) \bar{g}^{00}]^2} f'^2   \\   \nonumber
\Delta R_2&=& \frac{(1 - f) e^{2\sigma}}{1 + (1 -f) \bar{g}^{00}}
\Bigg[\left( \bar{g}^{\mu \nu} \Pi_{\mu \nu}^{(1)} - \bar{R}^{00} \right)
      + \frac{1 - f}{1 + (1 - f) \bar{g}^{00}}
        \left(\bar{g}^{\mu \nu} \Pi_{\mu \nu}^{(2)} - \bar{g}^{\mu 0}
        \bar{g}^{\nu 0} \Pi_{\mu \nu}^{(1)} \right)   \\   \nonumber
& &
\hspace{4.0cm}
- \left( \frac{1 - f}{1 + (1 - f) \bar{g}^{00}} \right)^2 
  \bar{g}^{\mu 0} \bar{g}^{\nu 0} \Pi_{\mu \nu}^{(2)} \Bigg].
\end{eqnarray}
In Eq.(\ref{remain1}) $\Pi_{\mu \nu}^{(1)}$ and $\Pi_{\mu \nu}^{(2)}$ are 
quantities dependent on the intrinsic geometry of brane world-volume as follows:
\begin{eqnarray}
\label{Pi-def}
\Pi_{\mu \nu}^{(1)}&\equiv& \nabla_{\nu} \omega^{\rho}_{\mu \rho} - 
\nabla_{\rho} \omega^{\rho}_{\mu \nu}    \\   \nonumber
\Pi_{\mu \nu}^{(2)}&\equiv& 2 \omega^{\rho}_{\mu \rho} \omega^{\sigma}_{\nu \sigma}
- \omega^{\rho}_{\mu \nu} \omega^{\sigma}_{\rho \sigma}
- \omega^{\sigma}_{\mu \rho} \omega^{\rho}_{\nu \sigma}
\end{eqnarray}
where $\nabla_{\mu}$ is a covariant derivative and
\begin{equation}
\label{omega-def}
\omega^{\mu}_{\nu \rho} = \frac{1}{2} \bar{g}^{\mu 0} \bar{g}^{\nu 0}
(\partial_{\nu} \bar{g}_{\rho \sigma} + \partial_{\rho} \bar{g}_{\nu \sigma}
- \partial_{\sigma} \bar{g}_{\nu \rho}).
\end{equation}
In fact, $\Delta R_2$ is not important in this letter because it goes to zero
at flat limit. Inserting $f(y) = 1 - \xi e^{4 k |y|}$ and $\sigma(y) = k |y|$ 
into Eq.(\ref{remain1}) one can show that $\Delta R_1$ is reduced to 
\begin{equation}
\label{dddr1}
\Delta R_1 = 16 k \left[ 1 - \frac{\xi}{2} \frac{2 + (1 + \xi) \bar{g}^{00}}
                                                {1 + \xi\bar{g}^{00}} \right]  \delta(y)
            - 20 k^2
           + \frac{4 k^2 \xi (1 + \bar{g}^{00}) e^{4 k |y|} [1 + 3 \xi \bar{g}^{00} e^{4 k |y|}]}
                  {(1 + \xi \bar{g}^{00} e^{4 k |y|})^2}.
\end{equation} 

Using Eq.(\ref{curvature}) and
\begin{equation}
\label{det}
\sqrt{-G} = \sqrt{-\bar{g}_4} e^{-4 \sigma}
\sqrt{\frac{1 + (1 - f) \bar{\mu}}{f}}
\end{equation}
where $\bar{g}_4 = det \bar{g}_{\mu \nu}$, $\bar{g}_3 = det \bar{g}_{ij}
(i,j=1,2,3)$, and $\bar{\mu} \equiv \bar{g}_3 / \bar{g}_4$, one can
calculate a $4d$ effective action whose form is
\begin{equation}
\label{effaction}
S_{eff} = \int d^4x \sqrt{-\bar{g}_4} {\cal L}_{eff}
\end{equation}
where
\begin{eqnarray}
\label{efflag}
{\cal L}_{eff}&=& \sqrt{\frac{1 + \bar{\mu} \xi}{1 - \xi}} 
			   (-v_b + {\cal L}_m)  \\  \nonumber
&+& \int dy e^{-4 \sigma} \sqrt{\frac{1 + (1 - f) \bar{\mu}}{f}}
\left[ -\Lambda + 2 M^3 (e^{2\sigma} \bar{R} + \Delta R_1 + \Delta R_2)\right].
\end{eqnarray}

As indicated earlier modern astronomical observations show that our universe
is flat in spite of its non-vanishing cosmological constant. In order to examine
whether or not it is realized at brane-world scenario we take a 
flat limit by choosing $\bar{\mu} = -1$,   
$\Delta R_1 = 16k(1 - \xi /2) \delta(y) - 20 k^2$, and $\Delta R_2 = 0$. Then 
one can show easily the effective Lagrangian contains a nonvanishing
cosmological constant
\begin{equation}
\label{efflag2}
{\cal L}_{eff} = 2 M_{pl}^2 \bar{R} + {\cal L}_m - \lambda_4
\end{equation}
where $M_{pl}^2  = M^3 / k$ and
\begin{equation}
\label{cosmoconst}
\lambda_4 = 24 M^3 k \left[ \frac{2}{3} \xi + \sqrt{1 - \xi} - 1 \right].
\end{equation}
The constant $\lambda_4$ is a quantity combined by the brane tension and the quantity 
arising from the contribution of the fifth dimension. 
In this sense it can be referred as an induced $4d$ cosmological constant.
Actually, however, in general there would be
a contribution from the matters ${\cal L}_m$ to the $4d$ cosmological constant.
For example, let us consider the scalar matter fields
\begin{equation}
\label{matter-def}
{\cal L}_m = - \frac{1}{2} \partial_{\mu} \phi^i \partial^{\mu} \phi^i - V(\phi).
\end{equation}
Due to homogenity and isotropy of our universe we assume the scalar fields
$\phi^i$ are dependent only on time, {\it i.e.} $\phi^i = \phi^i(t)$. Then, 
Eq.(\ref{emdefinition}) and Eq.(\ref{main}) yield
\begin{equation}
\label{matter-contri}
S_{ii} = {\cal L}_m = -24 M^3 k \left(1 - \frac{\xi}{3} - \sqrt{1 - \xi}\right).
\end{equation}
Thus the $4d$ effective Lagrangian (\ref{efflag2}) is reduced to 
${\cal L}_{eff} = 2 M_{pl}^2 \bar{R} - \Lambda_4$ where the $4d$ 
cosmological constant $\Lambda_4$ is 
\begin{equation}
\label{4d-cosmo-const}
\Lambda_4 = 8 M^3 k \xi.
\end{equation}
As expected $\Lambda_4$ becomes zero at zero temperature limit. At the 
cosmological temperature $3 {\bf K}$ with $M \approx k \approx 1 TeV$,
we obtain roughly $\Lambda_4 \approx (0.0004 eV)^4$ which is within the upper bound of the
experimantal value $(0.01 eV)^4$.

In summary, we examined the nonzero temperature effect in RS 
brane-world scenario. The assumptions that the RS bulk spacetime is 
two copies of the Schwarzschild-$AdS_5$ and finite temperature solution is 
a static limit of the RS cosmological solution enable us to derive a 
temperature dependence of the $4d$ cosmological constant at the effective 
action. The most interesting one is that the scenario presented in this
letter yields a small positive cosmological constant which is smaller than
the upper bound of the recent experimental value. 
This means that the modern astronomical observations can be realized within
the RS brane-world scenario by inclusion of temperature. It seems to be 
interesting to examine how the temperature effect modifies the Newton 
power law determined by the zero mode and higher Kaluza-Klein excitation
in the fluctuation spectrum, which will be discussed elsewhere.


\end{document}